\newfam\msbfam
\font\twlmsb=msbm10 at 12pt
\font\eightmsb=msbm10 at 8pt
\font\sixmsb=msbm10 at 6pt
\textfont\msbfam=\twlmsb
\scriptfont\msbfam=\eightmsb
\scriptscriptfont\msbfam=\sixmsb
\def\cj{\fam\msbfam}

\def\C{{\cj C}}

\def\R{{\cj R}}

\def\Z{{\cj Z}}

\centerline{\bf CHARGE CONJUGATION IN THE GALILEAN LIMIT}

\

\centerline{\bf M. Socolovsky} 

\

\centerline{\it  Instituto de Ciencias Nucleares, Universidad Nacional Aut\'onoma de M\'exico}
\centerline{\it Circuito Exterior, Ciudad Universitaria, 04510, M\'exico D. F., M\'exico} 

\

{\it Strictly working in the framework of the nonrelativistic quantum mechanics of a spin ${{1}\over{2}}$ particle coupled to an external electromagnetic field, we show, by explicit construction, the existence of a charge conjugation operator matrix which defines the corresponding antiparticle wave function and leads to the galilean and gauge invariant Schroedinger-Pauli equation satisfied by it. }

\


\

1. {\bf Introduction}

\

In a recent paper $^1$, Cabo {\it et al} showed the {\it existence} of the nonrelativistic limit $C_{nr}$ of the charge conjugation operation $C$ for the Dirac equation of a 4-spinor $\Psi=\pmatrix{\psi_1 \cr \psi_2 \cr \psi_3 \cr \psi_4 \cr}$ coupled to an external electromagnetic potential $(\phi, \vec{A})$. At low velocities of the Dirac particle with respect to the velocity of light in vacuum $c$, the ``large components'' $\psi=\pmatrix{\psi_1 \cr \psi_2 \cr}$ of $\Psi$ satisfy the Schroedinger-Pauli equation $^2$ $$i\hbar{{\partial}\over {\partial t}}\pmatrix{u \cr v \cr}={{1}\over{2m}}(-\nabla ^2 +{{q^2}\over{\hbar^2 c^2}}\vec{A}^2+{{iq}\over{\hbar c}}\nabla\cdot \vec{A}+{{2iq}\over{\hbar c}}\vec{A}\cdot \nabla-{{q}\over{\hbar c}}\vec{\sigma}\cdot\vec{B}+2mq\phi)\pmatrix{u \cr v \cr} \eqno{(1)}$$ where $q$ and $m$ are the electric charge and mass respectively, $\vec{\sigma}$ are the Pauli matrices, $\vec{B}=\nabla\times \vec{A}$ is the magnetic field and, at each space time point $\pmatrix{\vec{x} \cr t \cr}$, $\pmatrix{u \cr v \cr}\in \C^2$. The {\it charge conjugate} Pauli spinor $\psi_c$ representing spin ${{1}\over{2}}$ antiparticles (e.g. positrons) if $\psi$ represents spin ${{1}\over{2}}$ particles (e.g. electrons) is given by $$\psi_c=\pmatrix{-\bar{v} \cr \bar{u} \cr}=C_{nr}\pmatrix{u \cr v \cr} \eqno{(2)}$$ where $$C_{nr}=KM, \ \ \ M=\pmatrix{ 0 & -1 \cr 1 & 0 \cr} \eqno{(3)}$$ is the nonrelativistic limit of the charge conjugation matrix of the Dirac equation, which up to a sign is given by $^3$ $$C=i\gamma^2\gamma_0=\pmatrix{0 & 0 & 0 & -1 \cr 0 & 0 & 1 & 0 \cr 0 & -1 & 0 & 0 \cr 1 & 0 & 0 & 0}; \eqno{(4)}$$ $K$ is the complex conjugation antilinear and hemitian ($K^\dagger=K$) operation. $\psi_c$  satisfies the equation $$i\hbar {{\partial}\over{\partial t}}\pmatrix{-\bar{v} \cr \bar{u} \cr}={{1}\over {2m}}(\nabla^2-{{q^2}\over{\hbar^2 c^2}}\vec{A}^2+{{iq}\over{\hbar c}}\nabla \cdot \vec{A}+{{2iq}\over{\hbar c}}\vec{A}\cdot \nabla - {{q}\over {\hbar c}}\vec{\sigma}\cdot \vec{B}-2mq\phi)\pmatrix{-\bar{v} \cr \bar{u} \cr}. \eqno{(5)}$$ As it was proved in reference 1, both (1) and (5) are transformed into each other by the operator $C_{nr}$, thus reaffirming the {\it galilean character of the approximation $C_{nr}$ to $C$}. This is a non trivial result specially because of the general belief that charge conjugation is a symmetry that exists only in the relativistic regime. $^4$

\

In this note we discuss the previous result without appealing to the limiting process, namely, strictly working in the context of the galilean group, for simplicity of its connected component $G_0$, and of its universal covering group $\hat{G}_0$ (section 2). From the lagrangian density $\cal L$ for the equation (1), and using $C_{nr}$, we construct the lagrangian density ${\cal L}_c$ for equation (5), and prove the galilean invariance of these equations by proving this invariance for $\cal L$ and ${\cal L}_c$. We also verify the gauge invariance of ${\cal L}_c$ (section 3). 

\

2. {\bf Galilean group, its universal covering group, and spinors}

\

The connected component of the galilean group $G_0$ consists of the set of 4$\times$4 matrices $$g=\pmatrix{R&\vec{V}\cr 0&1\cr} \eqno{(6)}$$ with $R$ in the 3-dimensional rotation group $SO(3)$, boost velocity $\vec{V}$ in $\R^3$, composition law $$g_2g_1=\pmatrix{R_2 &\vec{V}_2 \cr 0 & 1 \cr}\pmatrix{R_1 & \vec{V}_1 \cr 0 &1\cr}=\pmatrix{R_2R_1 & \vec{V}_2+R_2\vec{V}_1 \cr 0 & 1 \cr}, \eqno{(6a)}$$ identity $$\pmatrix{I & 0 \cr 0 &1 \cr}, \ \ \ I=\pmatrix{1&0&0\cr 0&1&0 \cr 0&0&1}, \eqno{(6b)}$$ and inverse $$\pmatrix{R & \vec{V}\cr 0 &1}^{-1}=\pmatrix{R^{-1} & -R^{-1}\vec{V} \cr 0 & 1}. \eqno{(6c)}$$ $G_0$ is a non abelian, non compact, connected but non simply connected six dimensional Lie group; like the connected component of the Lorentz group, its topology is that of the cartesian product of the real projective space with ordinary 3-space {\it i.e.} of $\R P^3\times \R ^3$. The action of $G_0$ on spacetime is given by $$G_0\times \R ^4\to \R ^4, \ (g,\pmatrix{\vec{x}^\prime \cr t^\prime \cr})\mapsto \pmatrix{\vec{x}\cr t \cr}=g\pmatrix{\vec{x}^\prime \cr t^\prime \cr}=\pmatrix{R\vec{x}^\prime+\vec{V}t^\prime \cr t^\prime}. \eqno{(7)}$$ Since one has the action $$\mu:SO(3)\times \R ^3 \to \R ^3, \ (R,\vec{x})\mapsto R\vec{x}, \eqno{(8)}$$then $G_0$ is isomorphic to the semidirect sum $\R ^3 \times_{\mu}SO(3)$: $\pmatrix{R & \vec{V} \cr 0 & 1 \cr}\mapsto (\vec{V},R)$ with composition law $$(\vec{V}^\prime,R^\prime)(\vec{V},R)=(\vec{V}^\prime+R^\prime\vec{V},R^\prime R). \eqno{(8a)}$$

\ 

The {\it universal covering group} of $G_0$ is given by the $\Z_2$-bundle $$\Z_2\to \hat{G}_0 \buildrel {\Pi}\over \longrightarrow G_0 \eqno{(9)}$$ where $$\hat{G}_0=\{\hat{g}=\pmatrix{T & \vec{V} \cr 0 & 1 \cr}, \ T\in SU(2), \ \vec{V}\in \R ^3\}, \eqno{(9a)}$$ and $\Pi$ is the $2\to 1$ group homomorphism $$\Pi(\hat{g})=\pmatrix{\pi(T) & \vec{V} \cr 0 & 1 \cr} \eqno{(9b)}$$ with $\pi:SU(2) \to SO(3)$ the well known projection $$\pi\pmatrix{z & w \cr -\bar{w} & \bar{z} \cr}=\pmatrix{Rez^2-Rew^2 & Imz^2+Imw^2 & -2Rezw \cr -Imz^2+Imw^2 & Rez^2+Rew^2 & 2Imzw \cr 2Rez\bar{w} & 2 Imz\bar{w} & \vert z\vert ^2 -\vert w \vert ^2 \cr}. \eqno{(9c)}$$ $\hat{G}_0$ is simply connected and has the topology of $S^3\times \R ^3$. Since $SU(2)$ acts on $\R ^3$: $$\hat{\mu}:SU(2)\times \R ^3\to \R ^3, \ (T,\vec{V})\mapsto \pi(T)\vec{V},\eqno{(10)}$$ one has the group isomorphism $$\hat{G}_0\ni\pmatrix{T & \vec{V} \cr 0 & 1 \cr}\mapsto (\vec{V},T) \in \R ^3 \times _{\hat{\mu}}SU(2); \eqno{(11)}$$ the composition law in $\hat{G}_0$ is given by $$\pmatrix{T^\prime & \vec{V}^\prime \cr 0 & 1 \cr}\pmatrix{T & \vec{V} \cr 0 & 1 \cr}= \pmatrix{T^\prime T & \vec{V}^\prime+\pi(T^\prime)\vec{V} \cr 0 & 1 \cr}, \eqno{(12)}$$ while the identity and inverse are respectively given by $$\pmatrix{I & 0 \cr 0 & 1 \cr}, \ I=\pmatrix{1 & 0 \cr 0 & 1 \cr} \eqno{(12a)}$$ and $$\pmatrix{T & \vec{V} \cr 0 & 1 \cr}^{-1}=\pmatrix{T^{-1} & -\pi(T^{-1})\vec{V} \cr 0 & 1 \cr}. \eqno{(12b)}
$$ 

\

Turning back to physics, for each mass value $m>0$, $\hat{G}_0$ acts on the infinite dimensional Hilbert space ${\cal L}^2_1$ of continuously differentiable and square integrable $\C ^2$-valued functions $\pmatrix{u \cr v \cr}$ on $\R ^4$, the Schroedinger-Pauli spinors. This action is defined as follows: $^5$ $$\hat{\mu}_m:\hat{G}_0\times {\cal L}^2_1 \to {\cal L}^2_1, \ (\pmatrix{T & \vec{V} \cr 0 & 1 \cr},\pmatrix{u \cr v \cr})\mapsto \pmatrix{T & \vec{V} \cr 1 & 0 \cr}\cdot \pmatrix{u & \cr v \cr}: \R ^4 \to \C ^2,$$ $$\pmatrix{\vec{x} \cr t \cr}\mapsto \pmatrix{T & \vec{V} \cr 0 & 1 \cr}\cdot \pmatrix{u \cr v \cr}\pmatrix{\vec{x} \cr t \cr}= e^{{{-im}\over {\hbar}}(\vec{V}\cdot\vec{x}+{{1}\over{2}}\vert \vec{V}\vert ^2 t)}T\pmatrix{u(\pi(T)\vec{x}+\vec{V}t,t) \cr v(\pi(T)\vec{x}+\vec{V}t,t)}. \eqno{(13)}$$ $\hat{\mu}_m$ is equivalent to the {\it representation} $$\tilde{\hat{\mu}}_m:\hat{G}_0 \to End({\cal L}^2_1), \ \tilde{\hat{\mu}}_m(\hat{g})(\pmatrix{u \cr v & \cr})=\hat{g}\cdot \pmatrix{u \cr v \cr}. \eqno{(13.a)}$$ At each $t$ one has the inner product $$(\pmatrix{u_2 \cr v_2 \cr},\pmatrix{u_1 \cr v_1 \cr})(t)=\int d^3\vec{x}(\bar{u}_2(\vec{x},t)u_1(\vec{x},t)+\bar{v}_2(\vec{x},t)v_1(\vec{x},t)) \eqno{(14a)}$$ and the norm $$\vert \vert \pmatrix{u \cr v \cr} \vert \vert ^2 (t)=(\pmatrix{u \cr v \cr}, \pmatrix{u \cr v \cr})(t)=\int d^3\vec{x}(\vert u(\vec{x},t)\vert ^2 + \vert v(\vec{x},t)\vert ^2). \eqno{(14b)}$$ The galilean transformation of the charge conjugate spinor $\psi_c$ is given by $$\psi_c \mapsto \bar{\hat{g}}\cdot\psi_c, \ \pmatrix{\bar{T} & \vec{V} \cr 0 & 1 \cr}\cdot \pmatrix{-\bar{v} \cr \bar{u} \cr}\pmatrix{\vec{x} \cr t}=e^{{{im}\over{\hbar}}(\vec{V}\cdot\vec{x}+{{1}\over{2}}\vert \vec{V} \vert ^2t)}\bar{T}\pmatrix{-\bar{v}(\pi(\bar{T})\vec{x}+\vec{V}t,t) \cr \bar{u}(\pi(\bar{T})\vec{x}+\vec{V}t,t)}. \eqno{(15)}$$ Finally, the galilean transformations of the electromagnetic potential $(\phi, \vec{A})$ and the magnetic field $\vec{B}$ are $$\phi(\vec{x},t)=\phi^\prime(\vec{x}^\prime, t^\prime), \ \vec{A}(\vec{x},t)=R\vec{A}^\prime(\vec{x}^\prime,t^\prime), \ \vec{B}(\vec{x},t)=R\vec{B}^\prime(\vec{x}^\prime,t^\prime) \eqno{(16)}$$ with $\vec{x}=R\vec{x}^\prime+\vec{V}t^\prime$ and $t=t^\prime$. 

\

{\it Remark}: Representations associated with different values of the mass are inequivalent. $^6$

\

3. {\bf Lagrangian formulation and galilean and gauge invariances}

\

The Pauli equations (1) and (5) can be formulated within the lagrangian framework. The lagrangian for equation (1) is $${\cal L}={{i\hbar}\over {2}}((({{\partial}\over{\partial t}}-{{iq}\over{\hbar}}\phi)\psi^\dagger) \psi-\psi^\dagger ({{\partial}\over{\partial t}}+{{iq}\over{\hbar}}\phi)\psi)+{{\hbar ^2}\over{2m}}(\nabla+{{iq}\over{\hbar c}}\vec{A})\psi^\dagger\cdot(\nabla-{{iq}\over{\hbar c}}\vec{A})\psi-{{q\hbar}\over{2mc}}\psi^\dagger\vec{\sigma}\cdot\vec{B}\psi$$ $$={{i\hbar}\over{2}}(\dot{\psi}^\dagger\psi-\psi^\dagger\dot{\psi})+{{\hbar ^2}\over{2m}}\nabla\psi^\dagger\cdot\nabla\psi+{{q^2}\over{2mc^2}}\psi^\dagger\vert \vec{A}\vert ^2\psi+{{i\hbar q}\over{2mc}}(\psi^\dagger\vec{A}\cdot\nabla\psi-\nabla\psi^\dagger\cdot\vec{A}\psi)-{{q\hbar}\over{2mc}}\psi^\dagger\vec{\sigma}\cdot\vec{B}\psi+q\psi^\dagger\phi\psi, \eqno{(17)}$$ and equation (1) amounts to the variational equation $${{\delta}\over{\delta\psi^\dagger(\vec{x},t)}}S=0 \eqno{(18)}$$ where $S$ is the action $$S=\int dt\int d^3\vec{x}{\cal L}(\vec{x},t). \eqno{(19)}$$ Under the charge conjugation operation $${\cal L}\to {\cal L}_c=K{\cal L}=-{{i\hbar}\over {2}}((({{\partial}\over{\partial t}}+{{iq}\over{\hbar}}\phi)\psi^\dagger_c)\psi_c-\psi^\dagger_c ({{\partial}\over{\partial t}}-{{iq}\over{\hbar}}\phi)\psi_c)+{{\hbar ^2}\over{2m}}(\nabla-{{iq}\over{\hbar c}}\vec{A})\psi^\dagger_c\cdot(\nabla+{{iq}\over{\hbar c}}\vec{A})\psi_c+{{q\hbar}\over{2mc}}\psi^\dagger_c\vec{\sigma}\cdot\vec{B}\psi_c$$ 
$$=-{{i\hbar}\over{2}}(\dot{\psi}^\dagger_c\psi_c-\psi_c^\dagger\dot{\psi}_c)+{{\hbar ^2}\over{2m}}\nabla\psi^\dagger_c\cdot\nabla\psi_c+{{q^2}\over{2mc^2}}\psi^\dagger_c\vert \vec{A}\vert ^2\psi_c-{{i\hbar q}\over{2mc}}(\psi_c^\dagger\vec{A}\cdot\nabla\psi_c-\nabla\psi^\dagger_c\cdot\vec{A}\psi_c)+{{q\hbar}\over{2mc}}\psi^\dagger_c\vec{\sigma}\cdot\vec{B}\psi_c+q\psi^\dagger_c\phi\psi_c. \eqno{(20)}$$ 

To pass from (17) to (20), the identity $M^\dagger M=1$ is inserted at each term of (17), and the fact that $M\vec{\sigma}M^\dagger=M(\sigma_1,\sigma_2,\sigma_3)M^\dagger=(-\sigma_1,\sigma_2,-\sigma_3)$ is used; then the complex conjugation operation $K$ completes the transformation. 

\

The total action for the particle-antiparticle system is $$S_{tot}=S+S_c=\int dt \int d^3 \vec{x}({\cal L}(\vec{x},t)+{\cal L}_c(\vec{x},t)) \eqno{(21)}$$ and equation (5) is obtained from $S_{tot}$ or $S_c$ as $${{\delta}\over {\delta \psi^\dagger_c(\vec{x},t)}}S_{tot}={\delta \over {\delta \psi^\dagger_c(\vec{x},t)}}S_c=0. \eqno{(22)}$$

\

The lagrangian ${\cal L}$, and therefore the equation (1), are invariant under the galilean transformations (13), (15) and (16) for $\psi$, $\psi_c$, and $(\phi, \vec{A})$ and $\vec{B}$, respectively. To prove it, we use the facts that $\nabla=R^{-1}\nabla^\prime$ where $\nabla={{\partial}\over{\partial \vec{x}}}$ and $\nabla^\prime={{\partial}\over {\partial\vec{x}^\prime}}$, and ${{\partial} \over{\partial t}}={{\partial} \over{\partial t^\prime}}-R^{-1}\vec{V}\cdot\nabla^\prime$. If ${\cal L}^\prime$ and ${\cal L}^\prime_c$ are the transformed lagrangian densities for particles and antiparticles, then, from equation (20), $${\cal L}_c(\vec{x},t)=K{\cal L}(\vec{x},t)=K{\cal L}^\prime(\vec{x}^\prime,t^\prime)={\cal L}_c^\prime(\vec{x}^\prime,t^\prime) \eqno{(23)}$$ and therefore the galilean invariance of equation (5) is also proved. 

\

Finally, both ${\cal L}$ and ${\cal L}_c$, and therefore the equations (1) and (5), are gauge invariant under the transformations $\psi\to e^{i\Lambda}\psi$, $\psi_c\to e^{-i\Lambda}\psi_c$, $\phi \to \phi-{{\hbar}\over{q}}{{\partial}\over{\partial t}}\Lambda$ and $\vec{A}\to \vec{A}+{{\hbar c}\over{q}}\nabla\Lambda$, where $\Lambda$ is an arbitrary differentiable function of $(\vec{x},t)$.

\

{\bf Acknowledgement}

\

The author was partially supported by the project PAPIIT IN103505, DGAPA-UNAM, M\'exico. 

\

{\bf References}

\

1. A. Cabo, D. B. Cervantes, H. P\'erez Rojas and M. Socolovsky, Remark on charge conjugation in the nonrelativistic limit, {\it International Journal of Theoretical Physics} (to appear); arXiv: hep-th/0504223.

\

2. J. D. Bjorken, and S. D. Drell, {\it Relativistic Quantum Mechanics}, Mc Graw-Hill, New York (1964): p. 11.

\

3. M. Socolovsky, The CPT Group of the Dirac Field, {\it International Journal of Theoretical Physics}  {\bf 43}, 1941-1967 (2004); arXiv: math-ph/0404038.

\

4. V. B. Berestetskii, E. M. Lifshitz, and L. P. Pitaevskii, {\it Quantum Electrodynamics, Landau and Lifshitz Course of Theoretical Physics, Vol. 4}, 2nd. edition, Pergamon Press, Oxford (1982): p. 45.

\

5. J. A. de Azc\'arraga and J. M. Izquierdo, {\it Lie groups, Lie algebras, cohomology and some applications in physics}, Cambridge University Press, Cambridge (1995): p. 155.

\

6. S. Sternberg, {\it Group Theory and Physics}, Cambridge University Press, Cambridge (1994): p. 49.

\

\

\

\

\

e-mail: socolovs@nucleares.unam.mx, somi@uv.es

\

\end